%% file: main.tex
\documentclass[twocolumn]{aastex63}

\usepackage{amsmath}
\usepackage{amssymb}
\usepackage{amsthm}
\usepackage{graphicx}
\usepackage[colorinlistoftodos]{todonotes}
\usepackage{natbib}

\usepackage{lineno}
\usepackage{subfigure}
\usepackage{soul}
\usepackage{verbatim}

\setwatermarkfontsize{0.75in}
\definecolor{darkgreen}{rgb}{0.05,0.3,0.05}

\begin{document}

\vspace*{-\headsep}\vspace*{\headheight}
{\footnotesize \hfill FERMILAB-PUB-21-268-AE}\\
\vspace*{-\headsep}\vspace*{\headheight}
{\footnotesize \hfill DES-2021-0638}

\title{Expediting DECam Multimessenger Counterpart Searches with Convolutional Neural Networks}

\input{authors}

\begin{abstract}
Searches for counterparts to multimessenger events with optical imagers use difference imaging to detect new transient sources. 
However, even with existing artifact detection algorithms, this process simultaneously returns several classes of false positives: false detections from poor quality image subtractions, false detections from low signal-to-noise images, and detections of pre-existing variable sources. 
Currently, human visual inspection to remove the false positives is a central part of multimessenger follow-up observations, but when next generation gravitational wave and neutrino detectors come online and increase the rate of multimessenger events, the visual inspection process will be prohibitively expensive. 
We approach this problem with two convolutional neural networks operating on the difference imaging outputs. 
The first network focuses on removing false detections and demonstrates an accuracy of 92 percent on our dataset. 
The second network focuses on sorting all real detections by the probability of being a transient source within a host galaxy and distinguishes between various classes of images that previously required additional human inspection. 
We find the number of images requiring human inspection will decrease by a factor of 1.5 using our approach alone and a factor of 3.6 using our approach in combination with existing algorithms, facilitating rapid multimessenger counterpart identification by the astronomical community.
\end{abstract}

\keywords{Optical astronomy -- Machine learning -- Transient sources}


\section{Introduction}
\label{sec:introduction}
\input{introduction}

\section{Methods}
\label{sec:methods}
\input{methods}

\section{Results}
\label{sec:results}
\input{results}

\section{Discussion}
\label{sec:discussion}
\input{discussion}

\section{Conclusion}
\label{sec:conclusion}
\input{conclusion}

\section*{Acknowledgments}
\input{acknowledgements}

\software{
\texttt{ArtifactSpy} \citep[]{artifactspy},
\texttt{astropy} \citep[]{astropy},
\texttt{deeplenstronomy} \citep[]{deeplenstronomy}, 
\texttt{h5py} \citep[]{h5py},
\texttt{lenstronomy} \citep[]{lenstronomy},
\texttt{matplotlib} \citep[]{matplotlib},
\texttt{numpy} \citep[]{numpy}, 
\texttt{pandas} \citep[]{pandas},
\texttt{PlotNeuralNet} \citep[]{network_vis}, 
\texttt{PyTorch} \citep[]{pytorch},
\texttt{Scikit-Learn} \citep[]{sklearn}
}

\clearpage

\appendix
\numberwithin{figure}{section}
\numberwithin{table}{section}

\section{Label Correction with GradCAM}
\label{app:ml}
\input{appendix}

\bibliographystyle{yahapj_twoauthor_arxiv_amp}
\bibliography{main}

\end{document}

%% file: authors.tex
\correspondingauthor{Adam Shandonay and Robert Morgan}
\email{ashandonay@wisc.edu}
\email{robert.morgan@wisc.edu}

\author[0000-0002-2761-9319]{A.~Shandonay}
\affil{Physics Department, University of Wisconsin-Madison, 1150 University Avenue Madison, WI  53706, USA}

\author[0000-0002-7016-5471]{R.~Morgan}
\affil{Physics Department, University of Wisconsin-Madison, 1150 University Avenue Madison, WI  53706, USA}
\affil{Legacy Survey of Space and Time Corporation Data Science Fellowship Program, USA}

\author[0000-0001-8156-0429]{K.~Bechtol}
\affil{Physics Department, University of Wisconsin-Madison, 1150 University Avenue Madison, WI  53706, USA}
\affil{Legacy Survey of Space and Time, 933 North Cherry Avenue, Tucson, AZ 85721, USA}

\author[0000-0003-4383-2969]{C.~R.~Bom}
\affil{Centro Brasileiro de Pesquisas Físicas, Rua Dr. Xavier Sigaud 150, CEP 22290-180, Rio de Janeiro, RJ, Brazil}
\affil{Centro Federal de Educação Tecnológica Celso Suckow da Fonseca, Rodovia Mário Covas, Cep 23810-000, Itaguaí, RJ, Brazil}

\author[0000-0001-6706-8972]{B.~Nord}
\affil{Fermi National Accelerator Laboratory, P. O. Box 500, Batavia, IL 60510, USA}
\affil{Kavli Institute for Cosmological Physics, University of Chicago, Chicago, IL 60637, USA}

\author[0000-0001-9578-6322]{A.~Garcia}
\affil{Department of Astronomy, University of Michigan, Ann Arbor, MI 48109, USA}

\author[0000-0002-1448-219X]{B.~Henghes}
\affil{Department of Physics \& Astronomy, University College London, Gower Street, London, WC1 E 6BT, UK}

\author[0000-0001-6718-2978]{K.~Herner}
\affil{Fermi National Accelerator Laboratory, P. O. Box 500, Batavia, IL 60510, USA}

\author[0000-0002-0690-1737]{M.~Tabbutt}
\affil{Physics Department, University of Wisconsin-Madison, 1150 University Avenue Madison, WI  53706, USA}

\author[0000-0002-6011-0530]{A.~Palmese}
\affil{Fermi National Accelerator Laboratory, P. O. Box 500, Batavia, IL 60510, USA}
\affil{Kavli Institute for Cosmological Physics, University of Chicago, Chicago, IL 60637, USA}

\author[0000-0003-3402-6164]{L.~Santana-Silva}
\affil{NAT-Universidade Cruzeiro do Sul / Universidade Cidade de S{\~a}o Paulo, Rua Galv{\~a}o Bueno, 868, 01506-000, S{\~a}o Paulo, SP, Brazil}

\author[0000-0001-6082-8529]{M.~Soares-Santos}
\affil{Department of Physics, University of Michigan, Ann Arbor, MI 48109, USA}

\author[0000-0003-2524-5154]{M.~S.~S.~Gill}
\affil{SLAC National Accelerator Laboratory, Menlo Park, CA 94025, USA}

\author[0000-0002-9370-8360]{J.~Garc\'ia-Bellido}
\affil{Instituto de Fisica Teorica UAM/CSIC, Universidad Autonoma de Madrid, 28049 Madrid, Spain}

%% file: introduction.tex
\begin{figure*}
    \centering
     \includegraphics[width=0.45\textwidth]{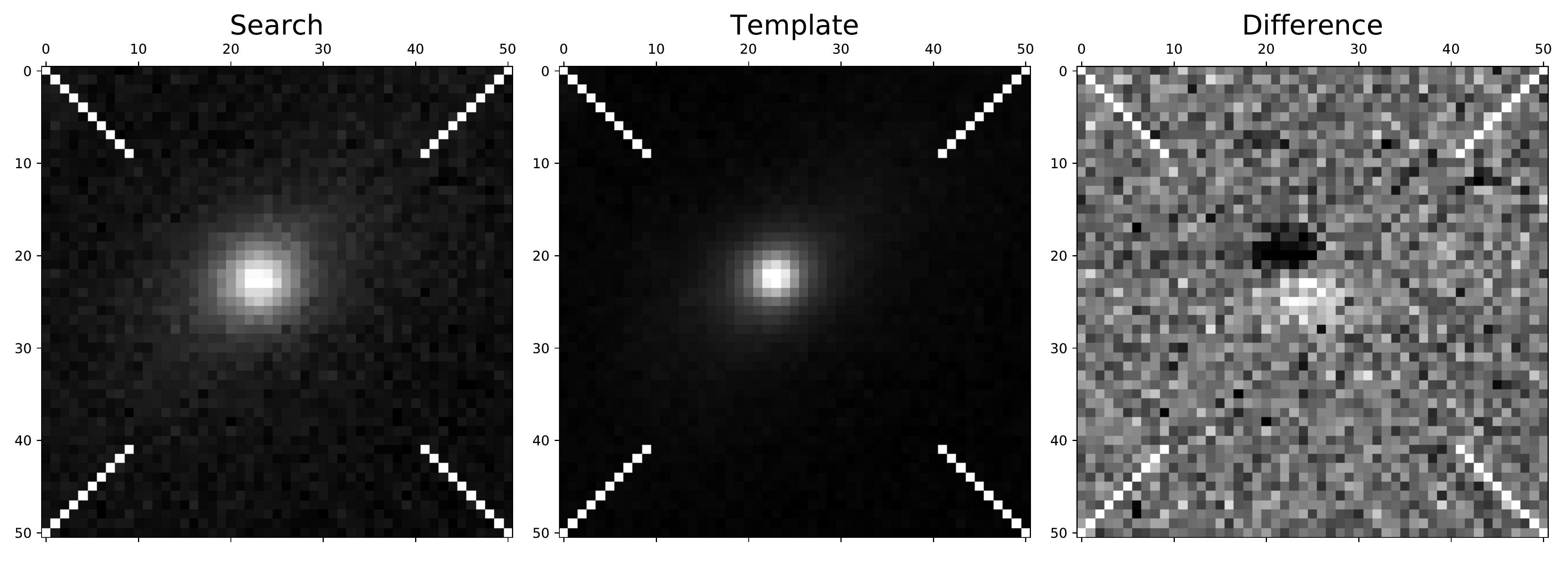}
     \hspace{1cm}
     \vspace{0.7cm}
     \includegraphics[width=0.45\textwidth]{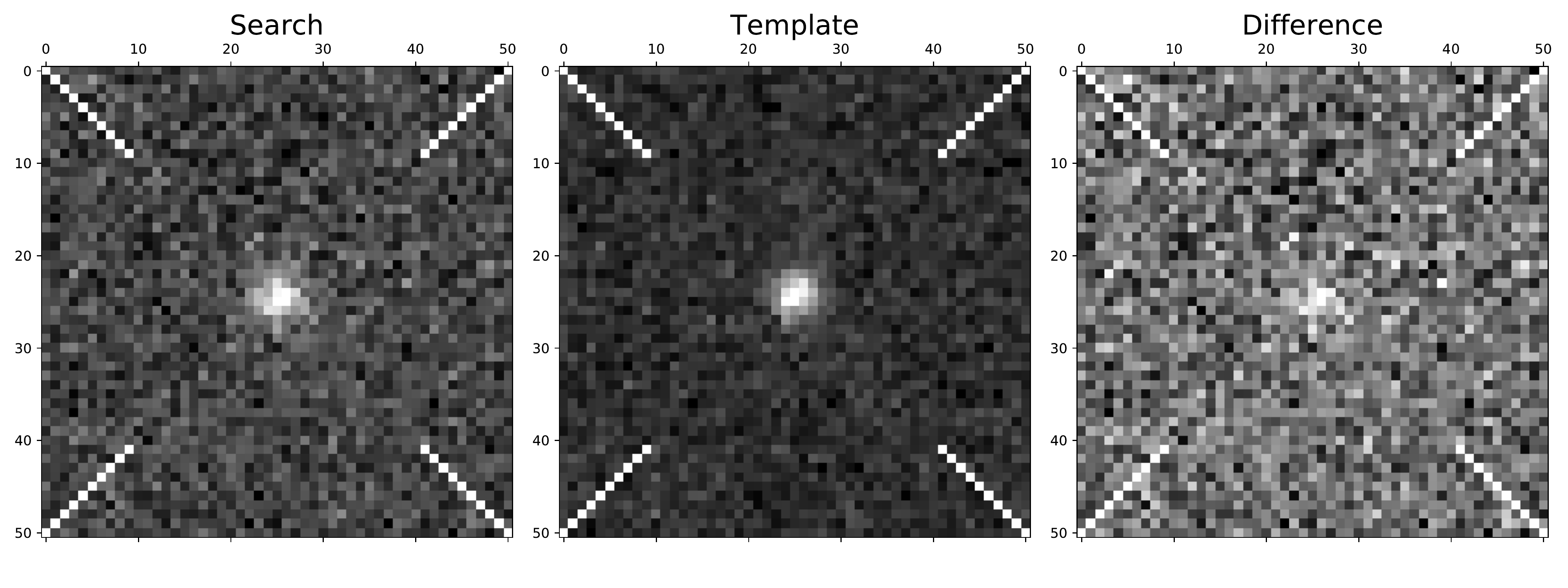}
     \includegraphics[width=0.45\textwidth]{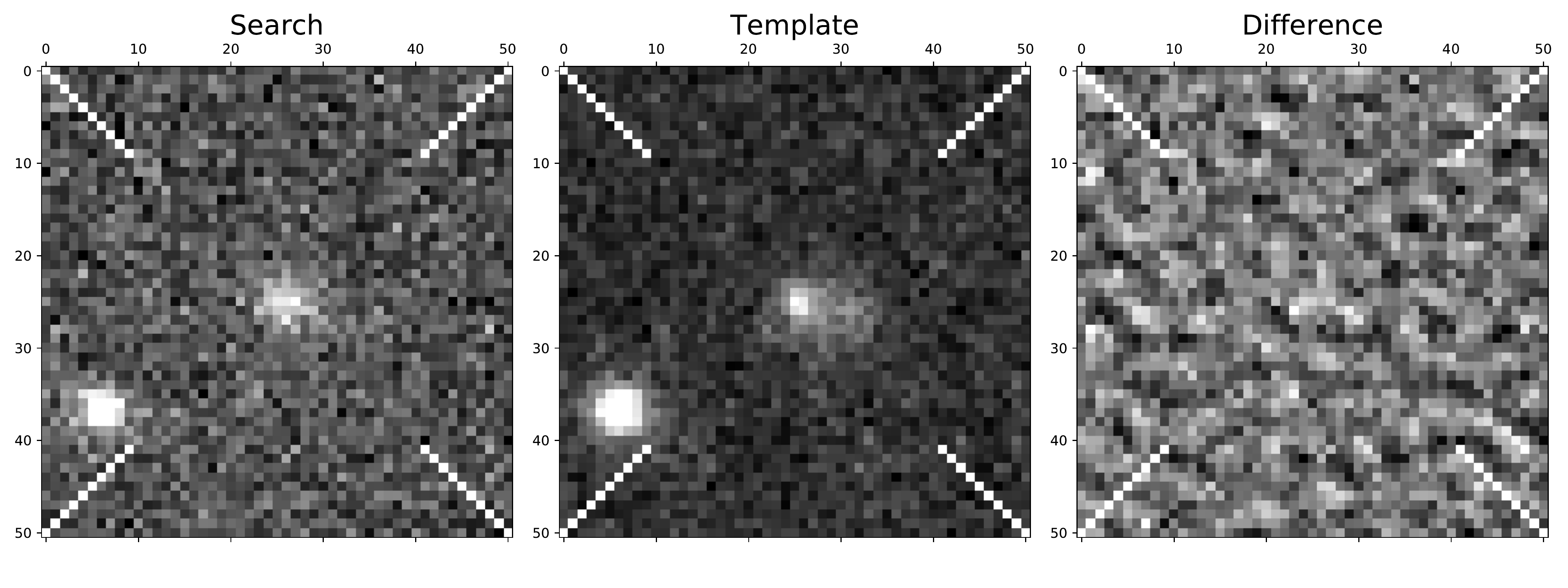}
     \hspace{1cm}
     \includegraphics[width=0.45\textwidth]{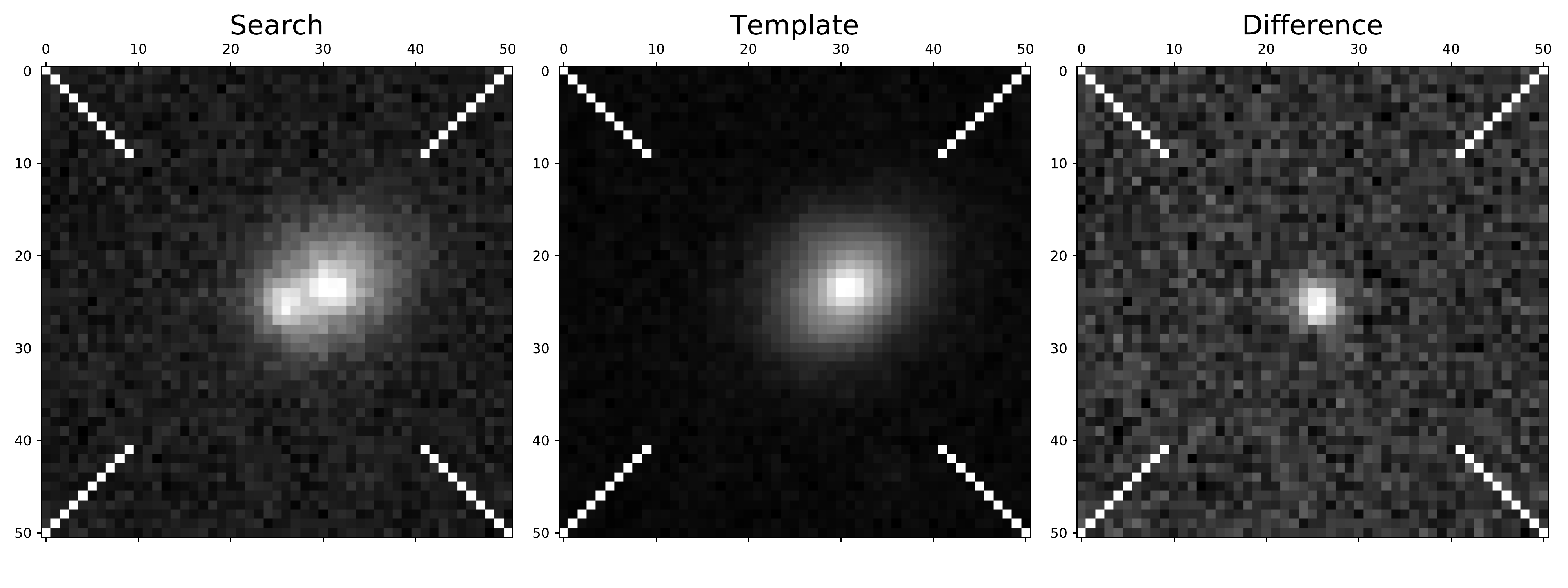}
      \caption{Examples of the four classes of difference imaging product in our dataset. The upper left panel shows a ``bad subtraction'' image.  The upper right panel shows a ``pre-existing point source'' image. The lower left panel shows a ``no obvious transient'' image. The lower right panel shows a ``transient + host'' image where a smaller, point-like object is visible within a host galaxy.} 
      \label{fig:examples}
\end{figure*}

Multimessenger astronomy utilizes the coordinated efforts of 
two or more types of detectors including electromagnetic, gravitational wave, and neutrino to gain increased understanding of astrophysical phenomena. 
Gravitational wave (GW) events detected by the Laser Interferometer Gravitational-Wave Observatory (LIGO) \citep{Aasi_2015} and Virgo \citep{Acernese_2014} or high-energy neutrinos detected by IceCube \citep{ACHTERBERG2006155} and ANTARES \citep{AGERON201111} may have electromagnetic counterparts that could offer insights for several fields of physics. 
For this analysis, we focus specifically on electromagnetic counterparts that emit in the optical wavelengths and are bright enough to be detected by existing instruments.
One such instrument, the Dark Energy Camera \citep[DECam][]{Flaugher_2015} mounted on the 4~m Victor M. Blanco Telescope at Cerro Tololo Inter-American Observatory in Chile has been a large contributor to the optical multimessenger follow-up community due to its $\sim3$ sq. deg. field of view and deep imaging capabilities.
Multiple observing teams have utilized DECam for multimessenger observations, and we expect DECam to continue to be a major multimessenger optical follow-up instrument into the Rubin Observatory Era.

The Dark Energy Survey Gravitational Wave (DESGW) team has developed a pipeline to efficiently search for new optical sources on large areas of sky during multimessenger follow-up campaigns with DECam \citep{HERNER2020100425}.
The search utilizes difference imaging \citep{Kessler_2015} to compare a recent (referred to as ``search'') image of the area in the sky associated with a gravitational wave or neutrino to a previous (referred to as ``template'') image of the same area taken at earlier times. 
After matching the point spread functions (PSFs) to the search and template images, the resulting pixel-by-pixel subtraction of the search and template images is called a ``difference image'' \citep{hotpants}. 
Once the extant light sources have been subtracted away, traditional astronomical image processing tools \citep{refId0} are applied to the difference images to identify new sources.
The goal of optical follow-up campaigns is then to scour the detections of new sources and rapidly communicate interesting objects to the astronomical community to facilitate multi-wavelength characterizations of any potential multimessenger counterparts.

In a typical counterpart search with DECam, the difference imaging pipeline produces $\sim10,000$ detections per 3~sq.~deq. field of view \citep[e.g. ][]{icecube_morgan, GW190814}.
These difference imaging datasets are dominated by artifacts --- several classes of visually obvious spurious detections described in Section \ref{sec:methods} --- , but both the size of the datasets and the need to find candidate objects quickly for spectroscopic characterization prohibits excluding these artifacts by visual inspection.
This problem can be remedied using automated artifact detection tools, of which there are several in use today \citep{DESAI201667}. 
Most of these automated tools employ machine learning at their cores, algorithmically developing selection rules for artifacts by analyzing large numbers of images.
The current tool used in the DESGW pipeline is \texttt{autoscan} \citep{Goldstein_2015}, which employs a Random Forest Classifier \citep{randomforest} to optimally weight hand-engineered features, such as signal-to-noise ratio, magnitude, etc.
An alternative machine learning approach, and the one we apply in this analysis, is to use deep learning, which learns selection rules directly from the data as opposed to from hand-engineered features.
In the case of images, the most common deep learning tools are convolutional neural networks (CNNs) \citep{cnn}. 

Applying CNNs to difference images has been done previously \citep{ztfrealbogus},  but this work is the first DECam-specific application in addition to a slight redefining of the classification problem.
In our analysis, we broaden the classes of objects to more accurately resemble a real multimessenger follow-up campaign; on top of filtering imperfect image subtractions from difference images, we also specifically look for the presence of a host galaxy and a new transient in the template and search images, respectively.
Our approach uses a pipeline of two CNNs trained on real DECam difference imaging data and image processing routines to algorithmically remove false positives from consideration for being a transient with a host galaxy.
The resulting output of our algorithm is a score for each image from 0 to 1 representing the probability of being a transient + host galaxy. 

Looking toward the expected increased multimessenger event rates of LIGO-Virgo-KAGRA fourth observing run \citep{O4} and IceCube-Gen2 \citep{gen2}, better performing tools to robustly and efficiently identify new transients in host galaxies will be a necessity.
This analysis presents one such tool that will enable DECam and the DESGW pipeline to efficiently perform multimessenger follow-up campaigns in the next era of multimessenger astronomy.
We organize the presentation as follows.
Section \ref{sec:methods} describes our training data, deep learning architecture, and training.
Section \ref{sec:results} presents the performance of each phase of our method.
We conclude Section \ref{sec:results} with end-to-end tests of our technique on several real DECam follow-up datasets.
Section \ref{sec:discussion} discusses the performance of our tool in the context of future multimessenger follow-up campaigns.
Lastly, we conclude in Section \ref{sec:conclusion}.

%% file: methods.tex
\subsection{Multimessenger Follow-Up Data}
When the difference imaging process works properly, the resulting difference image will contain transient objects that could potentially be the source of GWs or neutrinos. 
In practice, though, the rate of difference image detections is two orders of magnitude higher than the expected number of real transients.
Many false detections are caused by moving objects such as asteroids and satellites that can be ruled out easily using multiple observations, while other detections that cannot be eliminated simply are known as ``difference imaging artifacts''. 
The most common example of an artifact is known as a ``bad subtraction,'' where a slight misalignment or inaccurate determination of the PSFs between the search and template images creates adjacent under-subtracted and over-subtracted regions in the difference image, the first of which can be interpreted as a real object by Source Extractor \citep{refId0}. 
In the current DESGW pipeline, these bad subtractions are identified using a selection routine called \texttt{autoscan} \citep{Goldstein_2015} that assigns a score between 0 and 1 to each difference image with higher values corresponding to higher-quality detections. 
Most, but not all bad subtractions are removed with an \texttt{autoscan} threshold cut of 0.7, but this technique was not designed to remove other types of false detections.

Another type of false detection in counterpart searches is a pre-existing point source which is an object that was already visible in the template image, but produced a difference image due to changing brightness (e.g. variable Milky Way stars or astrophysical transients in the template image). 
Because these images contain real, astrophysical objects, they are often given high scores by artifact detection algorithms like \texttt{autoscan} even though they are of no interest in multimessenger astronomy.
The other common false positive that is not eliminated with \texttt{autoscan} is a marginal case where there is no obvious transient in the difference image.
These are images which seem to contain a host galaxy and a new object may appear in the search image, however, the resulting difference image is inconclusive. 
In most cases, this class contains galaxies with small variability in their centers rather than supernovae or kilonovae producing an obvious transient.

There are other less common types of false positives. 
For example, an asteroid detected in a previously empty patch of sky will appear to be a point-like source in the difference image and receive high \texttt{autoscan} scores because the lack of the presence of a host galaxy in the template image does not affect the scoring.
There are also cases where realizations of Poisson noise produce groups of pixels that could resemble an object in the search image, and in some cases groups of under-fluctuations in the template image which create the appearance of an object after subtraction.
We refer to this broad class as ``other artifact''.
The true positive case for multimessenger counterpart searches is when a distinguishable transient is visible only in the search image. 
The transient should exist within some host galaxy which will be present on both search and template images.
Figure \ref{fig:examples} displays examples of the four most common classes of object in our dataset.

\subsection{Data Collection}
\label{sec:data}

\begin{figure*}
\centering
\includegraphics[trim={0 2cm 0 0},clip,width=\textwidth]{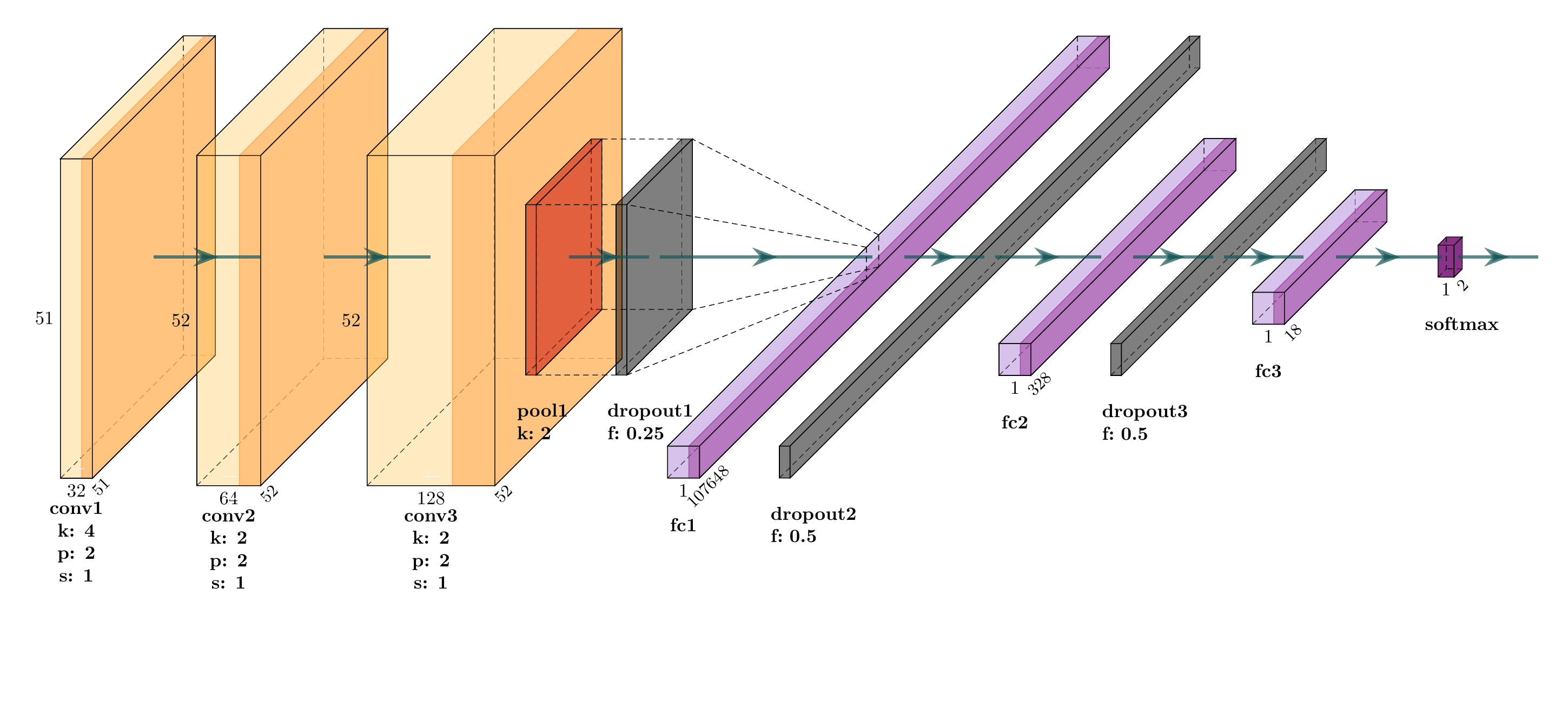}
\caption{The architecture of the neural networks utilized in our image classification algorithm, which extracts features using three convolutional layers (orange blocks) and classifies the images using fully-connected layers (purple) that weight and aggregate the extracted features. The height and depth of the blocks correspond to the image dimensions while the width corresponds to the number of channels (convolutional operators) applied. The shading at the right edge of the blocks indicates a ReLU activation function. The various letters mean the following: k = kernel size, p = padding, s = stride, f = dropout probability. The figure was produced using \texttt{PlotNeuralNet} \citep[]{network_vis}. \label{fig:network}}
\end{figure*}

Because we opted for a deep learning approach, we required a large, diverse set of real images with accurate labels. 
We utilized images of randomly sampled objects detected from applying the DESGW Search and Discovery Pipeline to DES wide-field data \citep[]{widediffimage} for our training data.
A team of six experts labeled the images corresponding to the five image types (``host + transient'', ``no obvious transient'', ``bad subtraction'', ``pre-existing point source'', ``other artifact'') using an interactive tool \citep[\texttt{ArtifactSpy};][]{artifactspy} that cycled the images across the team to ensure precise labeling.
When labeling the images, we rejected images whose detected transient had any masking over the transient, since the flux would be measured inaccurately.
We also rejected images where the detected transient could be matched to a high-confidence star in the DES Data Release 1 \citep{dr1} or GAIA Data Release 2 \citep{gaiadr2}, since these steps are routinely performed by the DESGW pipeline.

During the labeling process, it became clear that bad subtractions dominated the five image classes. 
To boost representation of the transient + host galaxy class, we obtained additional difference imaging data from three sources.
First, we supplemented the dataset with a population of the DES wide-field difference imaging data that was given an \texttt{autoscan} score of at least 0.9.
Second, we incorporated transient + host objects identified by human inspection during the DESGW follow-up observations of GW190814 \citep{GW190814} and GW200224 \citep[]{GCN_200224ca_1, GCN_200224ca_2}. 
Lastly, we simulated transient + host images using \texttt{deeplenstronomy} \citep[]{deeplenstronomy}. 
Importantly, the simulations are not run through the DESGW pipeline, so we are careful not to overload the training dataset with arbitrarily large amounts of simulated data.

From all the DES data and simulated data mentioned above, we construct two datasets.
In total, we obtained 1000 (640 real, 360 simulated) host + transient examples, 921 no obvious transient examples, 9,436 bad subtraction examples, 1050 pre-existing point source examples, and 731 other artifact examples.
Of the 640 real host + transient examples, 388 were collected from DES wide-field data, 214 were collected from the DESGW follow-up observations of GW200224, and 38 were collected from the DESGW follow-up of GW190814.
We also randomly down-sampled the bad subtraction class to 1,000 examples to create a less imbalanced classification problem.
Lastly, we applied rotations and mirroring to the images to increase the dataset size by a factor of 8.
The total collection of images was split into 90 percent training and 10 percent validation datasets.
Our overall approach is split into multiple parts, and in each part we use the training set to refine our algorithm and use the validation set to quantify its performance.

Furthermore, we also obtained additional real follow-up observation data which we only used for testing the entire approach in Section \ref{sec:real_test}.
Specifically, difference imaging samples from DECam follow-up observations of neutrino counterpart searches IC171106A \citep[]{icecube_morgan}, IC190331A, and IC201114A \citep{GCN_IC201114A_1} as well as GW counterpart searches GW190728 \citep{0728gcn} and GW190814.
These counterpart searches represent a variety of observing conditions and optical bandpasses.
We selected random samples of the difference images from these observations such that the size of the sample would produce a 68\% confidence level sampling error equal to 1\% of the entire population, indicating the sample size was large enough samples to be considered representative.
These datasets were kept separate from each other to assess the performance of our algorithm on standalone follow-up observations.

For reproduciblility, all images used for this analysis have been made publicly available \citep{zenodo-adam}.

\subsection{Algorithm Summary}
\label{sec:algorithm}

\begin{figure*}
    \centering
    \includegraphics[trim={1.5cm 1cm 1.5cm 1cm},width=0.45\linewidth]{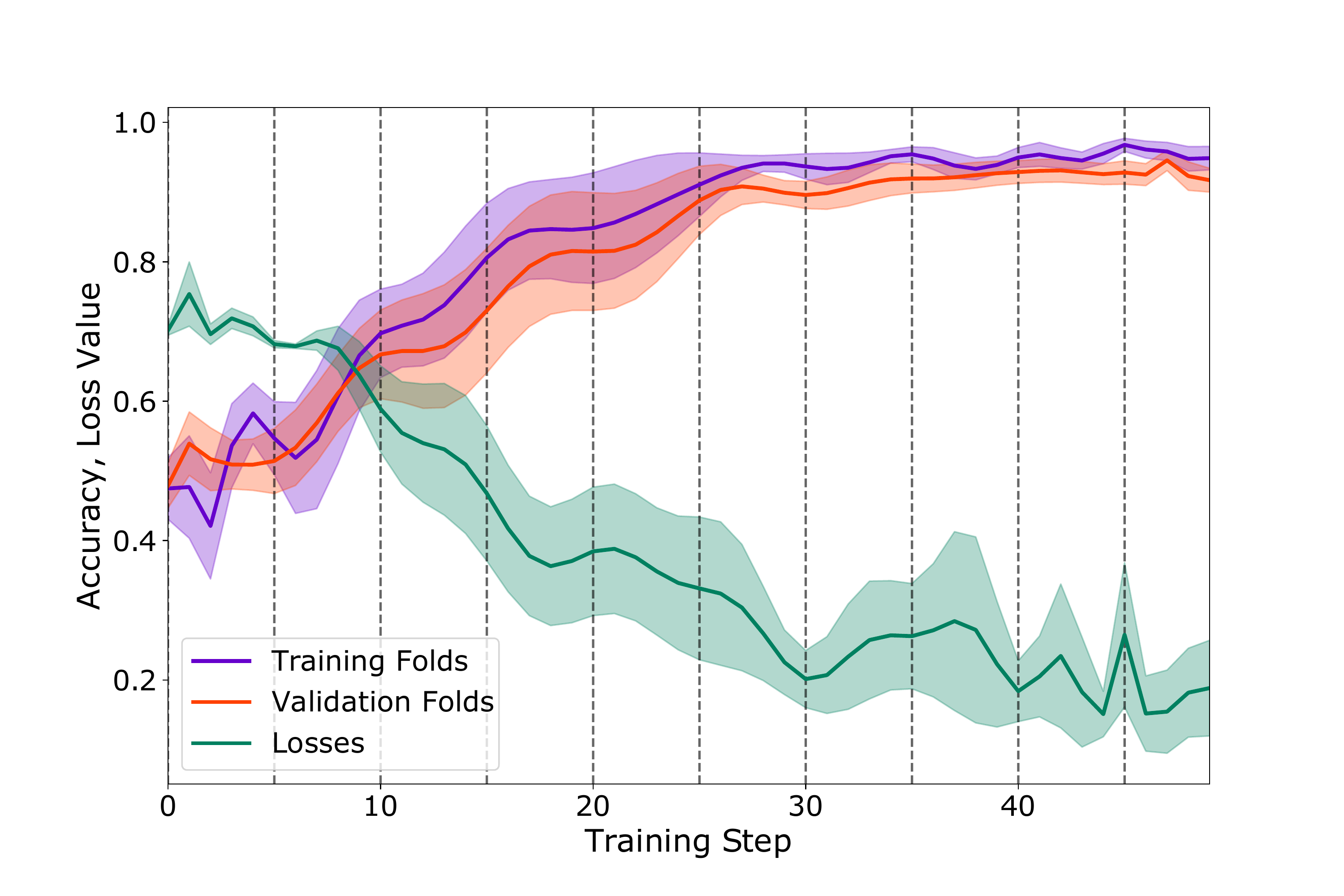}
    \includegraphics[trim={1.5cm 1cm 1.5cm 1cm},width=0.45\linewidth]{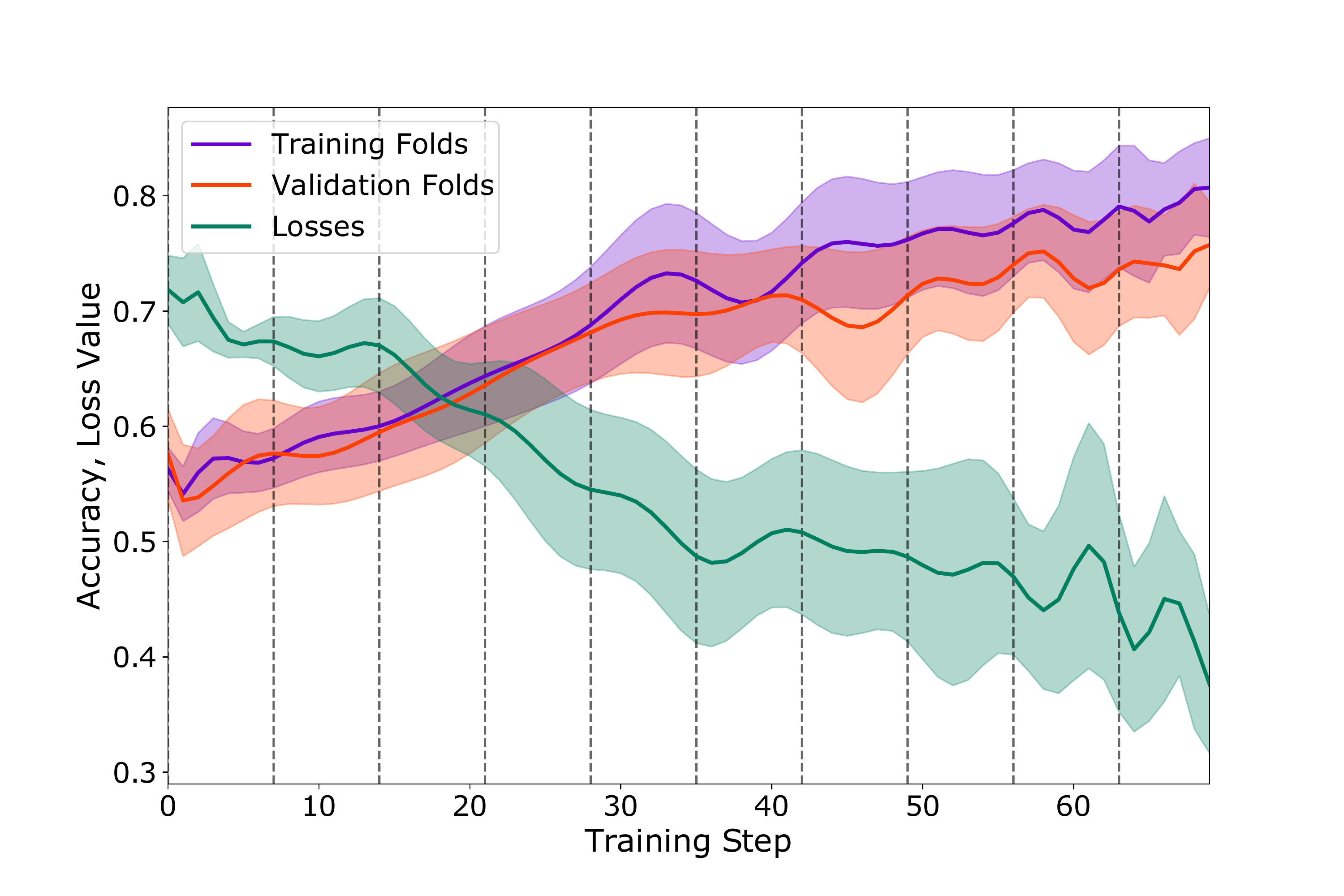}
    \caption{Performance on the training dataset from the first (left) and second (right) networks plotted using 10-fold cross validation. The standard deviation at each training step, corresponding to every tenth batch, is plotted using the shaded region around the mean value represented by the solid line. Additionally, the cross-entropy loss is plotted to show steadily decreasing values for both networks.}
    \label{fig:training}
\end{figure*}


The various classes of objects in difference imaging datasets present challenges for classification that demand a robust processing algorithm. 
Convolutional neural networks excel at image recognition \citep[]{lecun2015deep}, making them suitable candidates for improving image-based multimessenger counterpart searches.
However, we can simplify the classification process by applying high fidelity selection criteria to images before they are passed to a CNN. 
We apply two preprocessing filters to each set of search, template, and difference images designed to remove low quality or noisy images. 

The first preprocessing step performs an estimate of PSF flux at the center of the search image by subtracting the median value of the image, corresponding to the sky background, from each pixel of the image and weights the result by a Gaussian realization of the PSF from each image. 
We remove the images with a PSF flux below an empirically determined threshold.
Next, the remaining images go through a second preprocessing step which calculates the signal to noise ratio (SNR) from the \texttt{SExtractor} flux and flux error of the detection in the difference image and removes images with SNR below another empirically determined threshold. Both of these thresholds were determined by finding the strictest cut that would allow 99\% completeness of all classes not labeled other artifacts in the training dataset.
The choice of these threshold cuts are discussed in Section \ref{sec:results}.
Both of these steps aim to eliminate most of the noise detections in the other artifact class, since clear image features are necessary for a CNN to learn.

The vast majority of the remaining dataset examples are artifacts that are more difficult to remove with simple filters. 
We therefore develop deep learning tools to remove remaining false detections by learning features from the images.
First, a CNN is trained with the goal of exclusively identifying bad subtractions compared to the other classes. 
Singling out the bad subtractions ensures the highest level of accuracy when removing these artifacts from the dataset, as opposed to a multi-class classification scheme. 
The images classified as ``not bad subtraction'' pass the first CNN and move along a second CNN that scores images with a probability of being a real transient + host galaxy.
Finally, we use these probabilities from our method and the \texttt{autoscan} scores to fit a perceptron \citep{perceptron} that applies weights to both metrics and produces a combined score.
The remaining parts of this section describe each of the components of our approach in detail.

\subsection{Network Design}
\label{sec:network}

Convolutional Neural Networks are a particular type of deep learning tool that convolve a kernel of trainable weights with input data and learn feature maps that carry information regarding the identity of the objects in the input data. 
This kind of approach is particularly well-suited for computer vision tasks performing as the state-of-art in image classification because they optimize their feature maps through automated learning and back propagation of errors.
As a result, the input dataset can be diverse and still classified with high accuracy without the need for excessive amounts of training data.

The convolutional neural networks in our approach were developed using the \texttt{PyTorch} library \citep[]{pytorch}. 
Both networks have the same structure with three convolutional layers,  a max pooling layer, three dropout layers, and three fully-connected layers. 
The network architecture has two main components: a feature extraction step where the convolutional layers locate edges and shapes within the images, and a classification step where the fully-connected layers weight the extracted features and reduce them to classifications.
Figure \ref{fig:network} illustrates the flow of information through the layers of the network, as well as the hyperparameter settings utilized in our analysis.

The search, template, and difference images are simultaneously fed into the network through independent channels, similar to how the red, green, and blue image arrays are fed into CNNs in traditional image analyses.
The first convolutional layer applies 32 convolutional filters with size 4 px and stride 1 px to the 51 px by 51 px search, template, and difference images.
Following traditional approaches \citep{resnet, inception}, the number of convolutional filters is doubled with each additional layer, and we reduce the size of the kernel by a factor of 2 with each additional layer to focus on smaller scale image features.
We maintain a stride of 1 px for each layer, which prevents downsampling of the image representation size during the convolutions.
After all three convolutional layers, we apply max pooling with a kernel size of 2 px to magnify the importance of the main features found by the convolutional operations.
Lastly, we flatten the image representation and apply a series of fully connected layers to produce classifications from the image representations.
With each fully-connected layer, we reduce the size of the representation to approximately its square root until reaching a size of 2 (a probability for each group in our binary classification scheme).
Importantly, before each fully connected layer, we utilize a dropout layer to prevent overfitting.
Each of the hyperparameter values above was optimized, as described in Section \ref{sec:tuning}.
For reproduciblility, all neural network code used for this analysis has been made publicly available \citep{zenodo-adam}.


\begin{figure*}
\centering
  \includegraphics[width=0.49\textwidth]{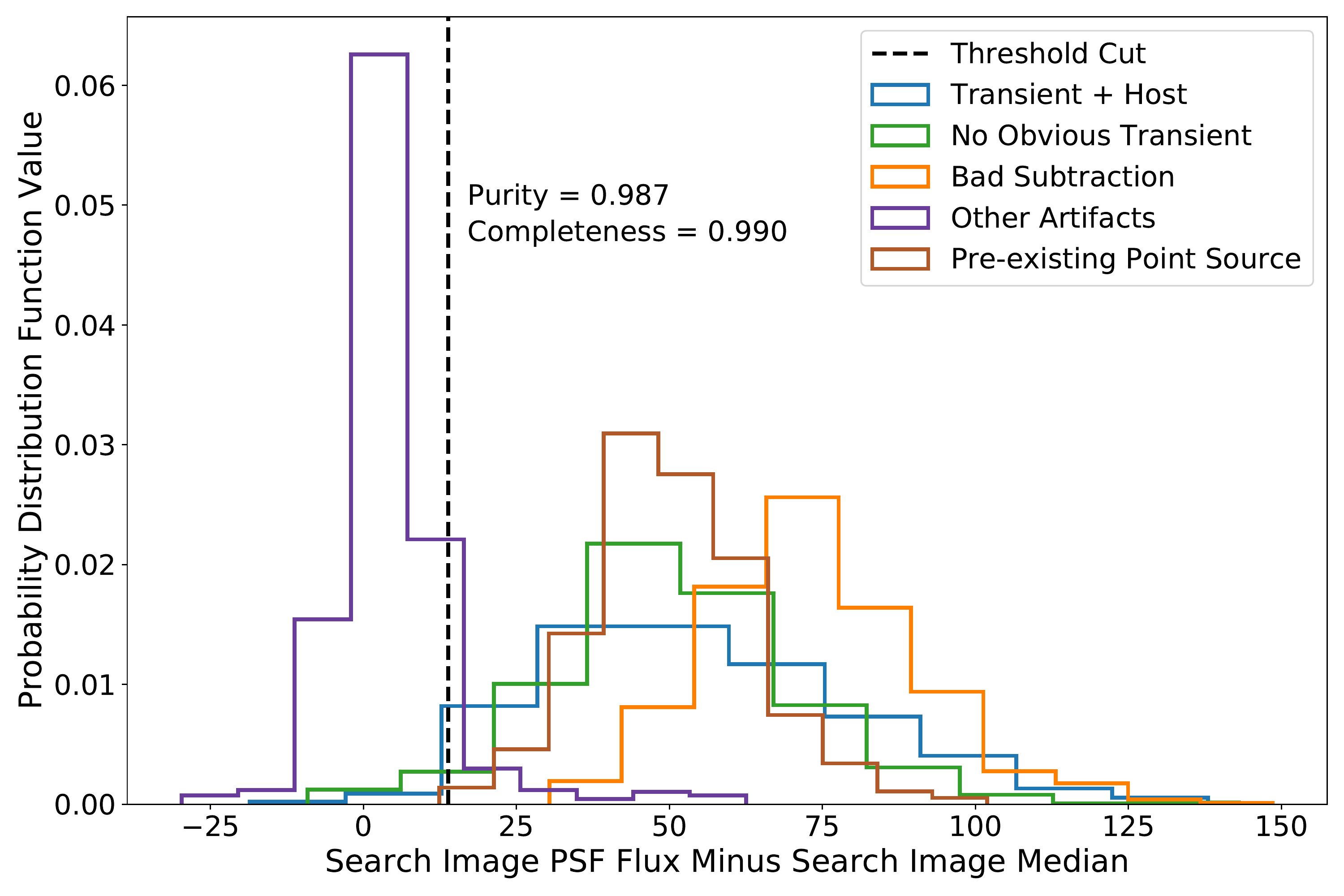}
  \includegraphics[width=0.49\textwidth]{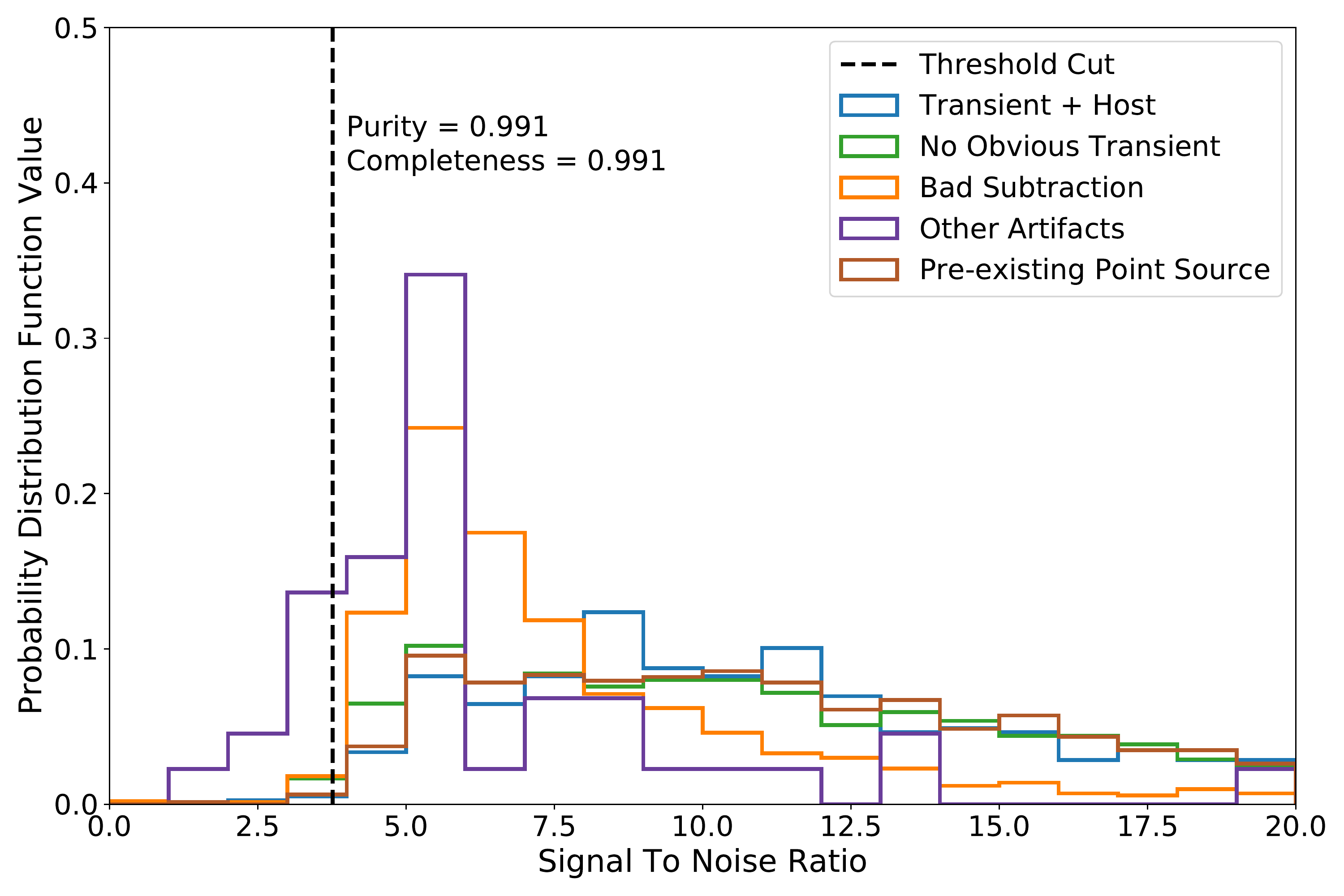}
  \caption{Left: Histogram of weighted flux values from first preprocessing step. Right: Histogram of signal to noise ratio values from second preprocessing step. Both panels show the validation dataset. The purity and completeness was calculated by identifying the other artifacts class as negative and all other classes as positive because we only intended to remove those images in this step. \label{fig:preprocessing}}
\end{figure*}

\subsection{Hyperparameter Optimization}
\label{sec:tuning}

The convolutional layers of the neural networks can be modified by adjusting hyperparameters that determine how the CNN processes each image. 
All of the hyperparameters were chosen by varying the values and testing the CNN for performance improvements until the highest accuracy was achieved.
For this optimization, we trained our network architecture on the training dataset and predicted the labels of the objects in the validation dataset.
We perform 5-way classification, since our goal was to determine the network hyperparameters best for picking out the features for all the classifications we desired to make.

The adjustment of most hyperparameters within reasonable ranges did not produce significant effects.
Specifically, varying the stride of the convolutional layers from 1 pixel to 2 pixels, varying the dropout fractions from 0.1 to 0.5, and varying the initial number of convolutional filters from 8 to 32 did not result in changes to the accuracy on the validation set above the typical levels of accuracy fluctuations.
We did not perform a hyperparameter search of learning algorithm parameters such as the learning rate, loss function, batch size, and optimizer because the accuracy for our fiducial experimental configuration was satisfactory for the goals our this analysis.

Two hyperparameters significantly affected the performance: the kernel size of the first convolutional layer and the number of convolutional layers.
The kernel size is the dimension of a matrix of weights that slides across the image and detects features by multiplying each matrix element by the corresponding pixel on the image. 
The optimal kernel size depends on the size of features in the image, so certain values will give better performance. 
In a test of 1 pixel to 16 pixels, a kernel size of 4 pixels was found to be optimal, but similarly high performance was observed for kernel sizes between 1 pixel to 8 pixels, corresponding to an upper limit of roughly 2 arcseconds (1 DECam pixel = 0.263 arcsec). 
We interpret this result as the network's ability to detect the presence of host galaxies approximately 2 arcseconds in size.
We therefore utilized a kernel size of 4 pixels as our largest kernel and decreased the kernel size in the deeper layers of the network to find smaller features such as point source transient objects.
When trying to identify smaller-scale features, we also found that a third convolutional layer was beneficial when using initial kernels sizes of 4 pixels and greater, likely due to the relative scales of features being identified by the network.

\subsection{Training}
\label{sec:training}

After optimizing the architecture of our network, we opted to reformulate the classification problem based on our observations.
In the five-way classification scheme, the majority of the confusion was between the host + transient class and the bad subtraction class.
In an attempt to prevent this confusion from hindering performance in a real multimessenger follow-up campaign dataset --- where bad subtractions often occur in large numbers --- we chose to focus on removing the bad subtractions with a standalone network.
For this first standalone network, the training was performed using only the transient + host class to represent a positive result and the bad subtractions to represent a negative result.
For the remainder of the analysis, the goal became to identify the host + transient class from anything not labeled as a bad subtraction by the first network.
Therefore, a second network was utilized to perform two-way classification of host + transient examples versus non-host + transient examples.

We characterized the performance of the networks during training by employing a 10-fold cross validation of the training dataset.
The networks were trained for 50 epochs and 70 epochs respectively.
Network 2 was allowed to train longer due to higher observed fluctuations in the loss of the validation folds of the training dataset.
Figure \ref{fig:training} shows the performance of the networks during 10-fold training on just the training dataset, while Sections \ref{sec:firstnet} and \ref{sec:secondnet} present the measured performance of the trained networks on the validation dataset as a result of the analysis.
From Figure \ref{fig:training}, we observe high training accuracies, but more importantly, we do not observe large amounts of overfitting.



%% file: results.tex
After using the training dataset to select the preprocessing filters' thresholds and train the two CNNs, we measure the performance of our algorithm in several tests using the validation dataset and real follow-up observations. 
In this section we present several performance metrics.
We measure the efficacies of the preprocessing filters and CNNs on the validation dataset.
We then compare the output probabilities of our second CNN, the goal of which was to sort the remaining images by the probability of being a transient in a host galaxy, to the \texttt{autoscan} scores to demonstrate improvement over tools currently in use.
We quantify the selection function of our approach by measuring the purity, completeness, and false positive rate as functions of physical properties.
Lastly, we apply our full algorithm to unseen data from real follow-up campaigns to quantify the factor of improvement in optical counterpart finding efficiency.

\subsection{Preprocessing}
\label{sec:preprocessing}
The effectiveness of the preprocessing filters are displayed in Figure \ref{fig:preprocessing} using histograms. 
The output from the first step of calculating a weighted value for the flux of each image after subtracting the background is shown in the left panel of Figure \ref{fig:preprocessing}. 
The threshold was set by requiring a completeness value of 0.99 to ensure a high number of positive images, represented by all classes not defined as ``Other Artifacts'', are included. 
The resulting purity value was 0.987 at a threshold set to 13.86 such that nearly all passed images are true positives. 
The output from the second step of calculating a signal to noise ratio of each image is shown in the right panel of Figure \ref{fig:preprocessing}.
The threshold was again set by requiring a completeness of 0.99 such that a high number of the positive images are included. 
The resulting purity value of 0.991 at a threshold set to 3.76 such that nearly all passed images are true positives.
Together, these two preprocessing filters remove large fractions of the ``Other Artifacts'' class, which simplifies the classification problem for the deep learning parts of our approach.



\subsection{First Network}
\label{sec:firstnet}

\begin{figure*}
\centering
\includegraphics[trim={0 0 1cm 0cm},clip, width=0.4\linewidth]{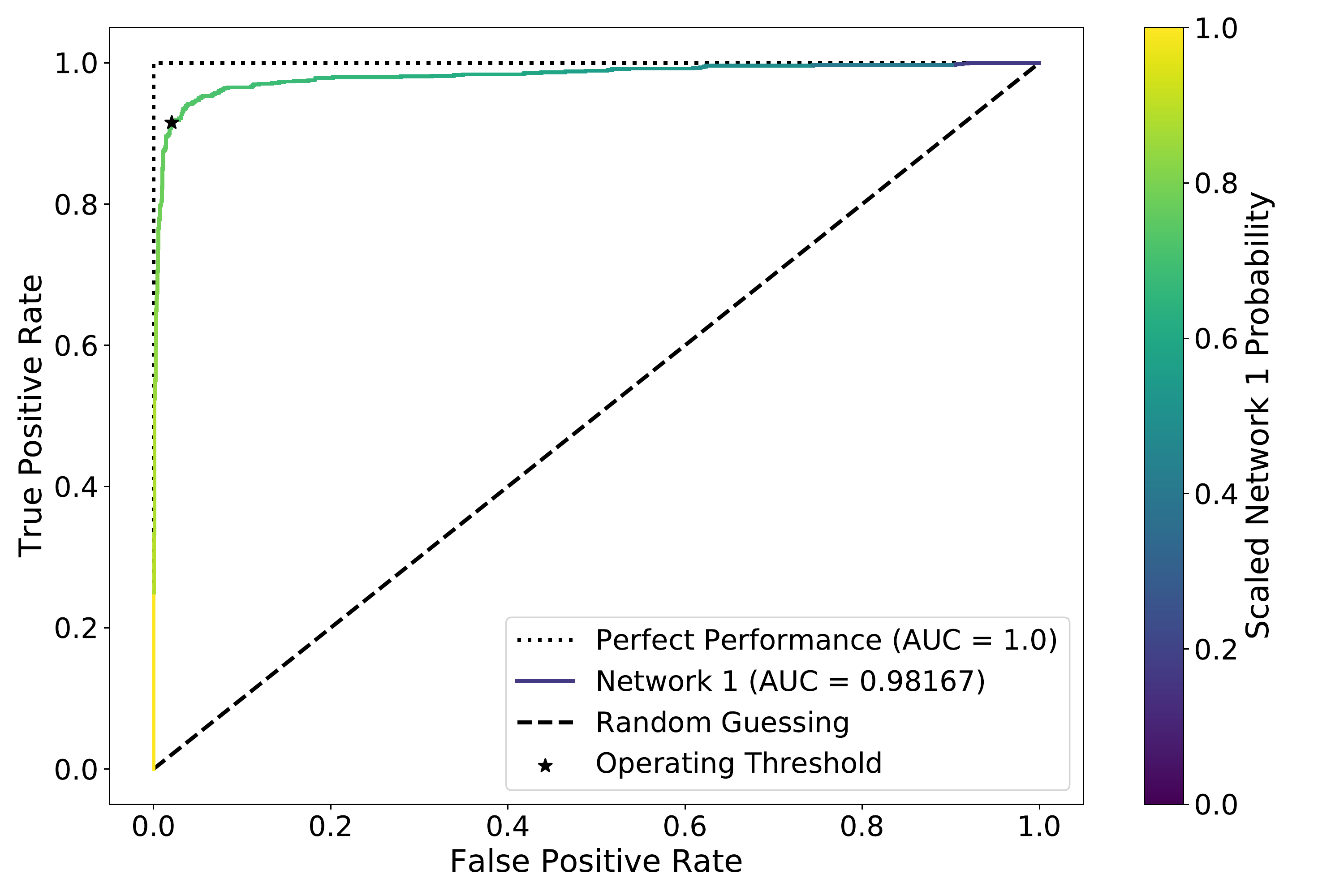}
\includegraphics[trim={0 3.5cm 0 0cm},clip, width=0.59\linewidth]{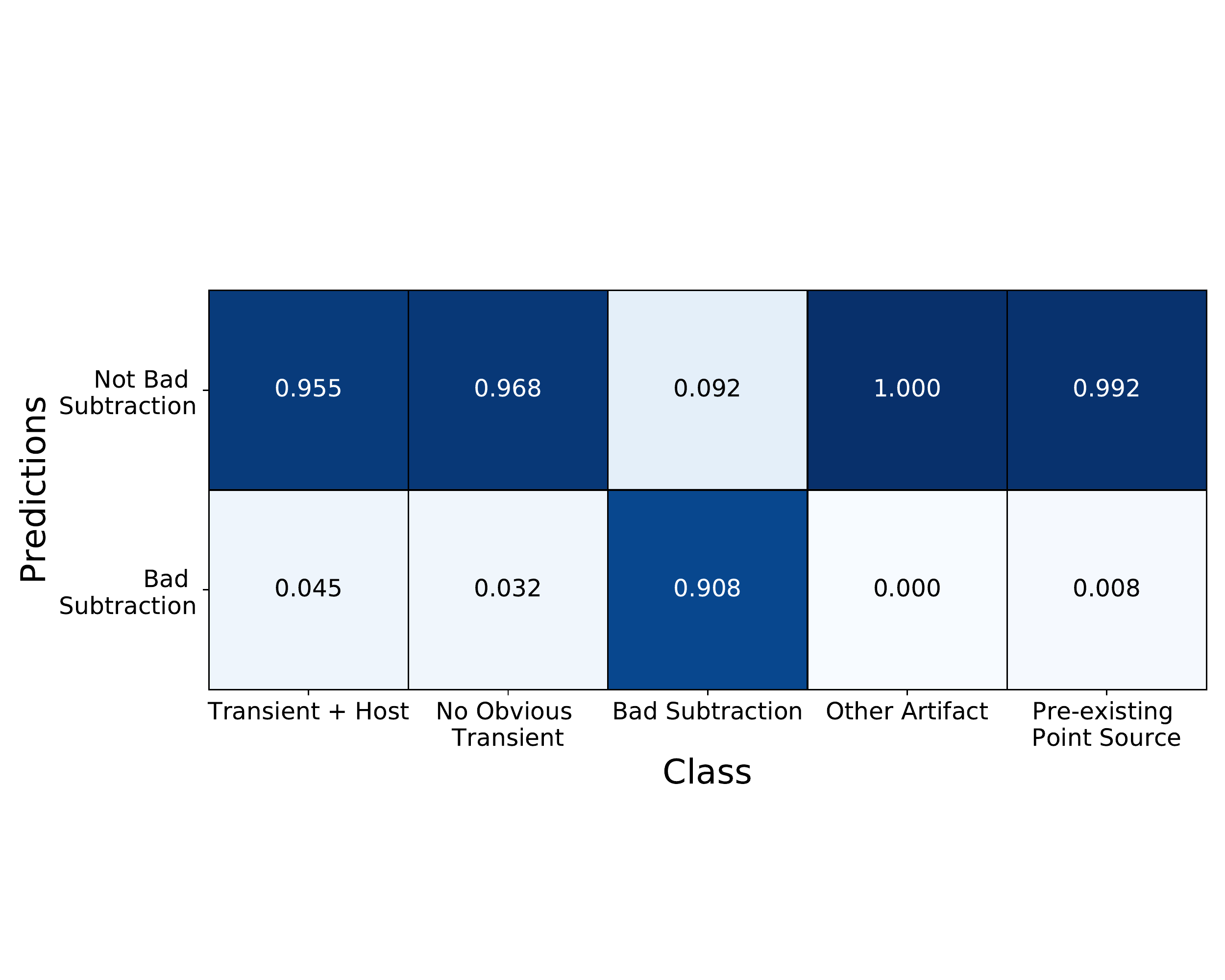}
\caption{Left: ROC curve of the first CNN's output. Right: Confusion matrix of first CNN predictions. Both panels display results from apply the trained networks to the validation dataset. \label{fig:network1_performance}}
\end{figure*}



The first CNN in the algorithm is applied to the images that passed the preprocessing filters.
Using the validation dataset, the output probabilities were used to construct a Receiver Operating Characteristic (ROC) curve to evaluate the ideal operating threshold. 
The curve is plotted in Figure \ref{fig:network1_performance}. 
The area under the curve (AUC) of the first network is 0.982 which is near perfect performance. 
The operating threshold for determining positive and negative results was calculated by maximizing the F1 score which is the harmonic mean of the purity and completeness.
Using this threshold, the false positive rate was 3\% and the true positive rate was 92\%. 
This means that only 3\% of the non-bad subtraction images were incorrectly called bad subtraction and will not be passed to the second CNN. 
A confusion matrix of the binary predictions using the chosen operating threshold is shown in Figure \ref{fig:network1_performance}. 
All non-bad subtraction classes demonstrated an accuracy greater than 95\% which was the intended result of using a straightforward binary classification.

\subsection{Second Network}
\label{sec:secondnet}

The final step in the classification algorithm is applied to the images which were not identified as bad subtractions by the first CNN or removed as other artifacts by the preprocessing.
The second network is used to sort the real detections, the images excluding bad subtractions and other artifacts, by their probability of being a real transient source with a host galaxy. 
The probabilities are shown in blue for the three primary remaining classes in 
Figure \ref{fig:autoscan-compare}. 
The majority of the transient + host class falls above a fiducial 0.5 probability threshold of being labeled correctly. 
Nearly all of the pre-existing point sources are correctly labeled below the 0.5 probability threshold indicating a clear distinction between these two classes. 
A large portion of the ``No Obvious Transient'' class falls at high probabilities which shows the network's ability to identify useful images that are not distinguishable with visual inspection.

\subsection{Comparison to \texttt{autoscan}}
\label{sec:algorithmresults}

\begin{figure}
  \includegraphics[width=0.45\textwidth]{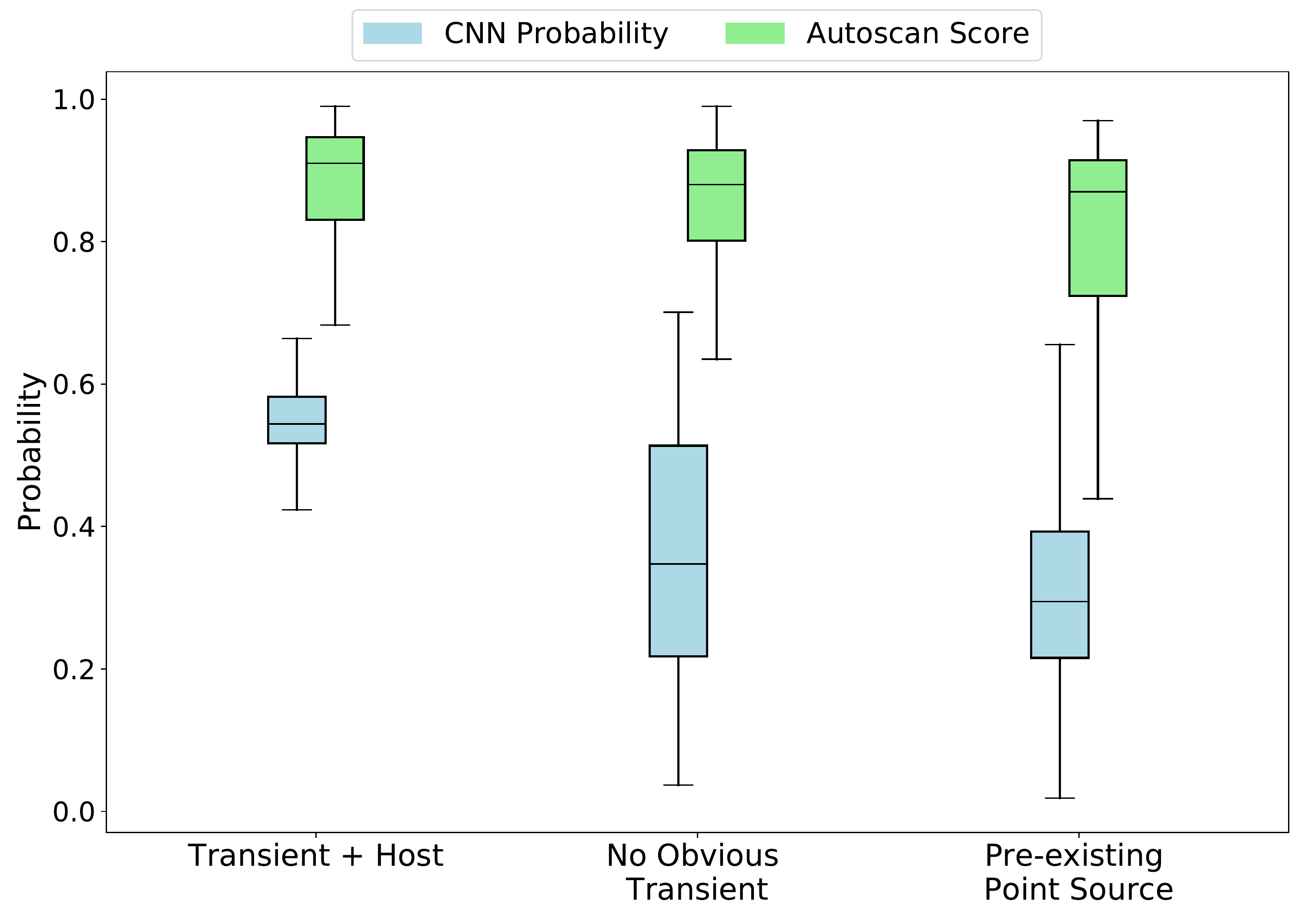}
  \caption{Final probabilities of being a transient + host galaxy compared to \texttt{autoscan} probabilities. The boxes extend from the lower to upper quartile of each group with the median marked by the line. \label{fig:autoscan-compare}}
\end{figure}

Multiple tests were conducted to examine whether the algorithm has the intended result of expediting multimessenger counterpart searches. 
The first such test was comparing the output probabilities from the second network with the scores given to images by \texttt{autoscan}.
\texttt{autoscan} is the current method for identifying difference image artifacts mainly focused on removing bad subtractions, but it was not designed to distinguish other types of false detections. 
The comparison is shown in Figure \ref{fig:autoscan-compare}.

The biggest takeaway from Figure \ref{fig:autoscan-compare} is that \texttt{autoscan}'s score shown in green has similar median values around 0.9 and similar upper and lower quartiles ranges, which means the not obvious transient class and pre-existing point source class are not being distinguished from the transient + host galaxy class. 
With our image classification algorithm, there is a significant difference in the probabilities assigned to transient + host compared to the other classes with higher median values. 
This means the output probabilities can be used as improved indicators of real transients.

\begin{figure*}
    \centering
    \includegraphics[width=0.49\linewidth]{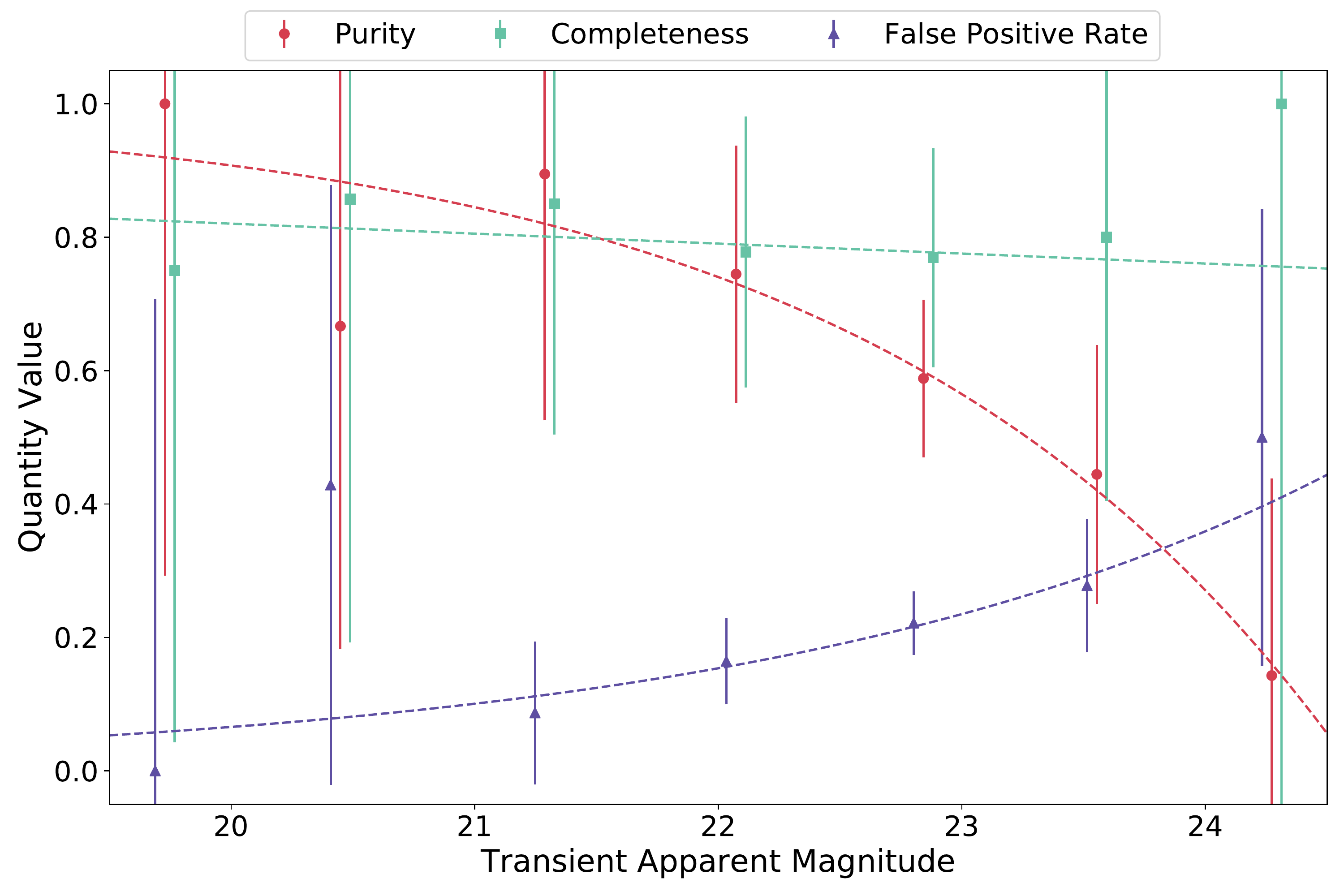}
    \hspace{0.008\linewidth}
    \includegraphics[width=0.49\linewidth]{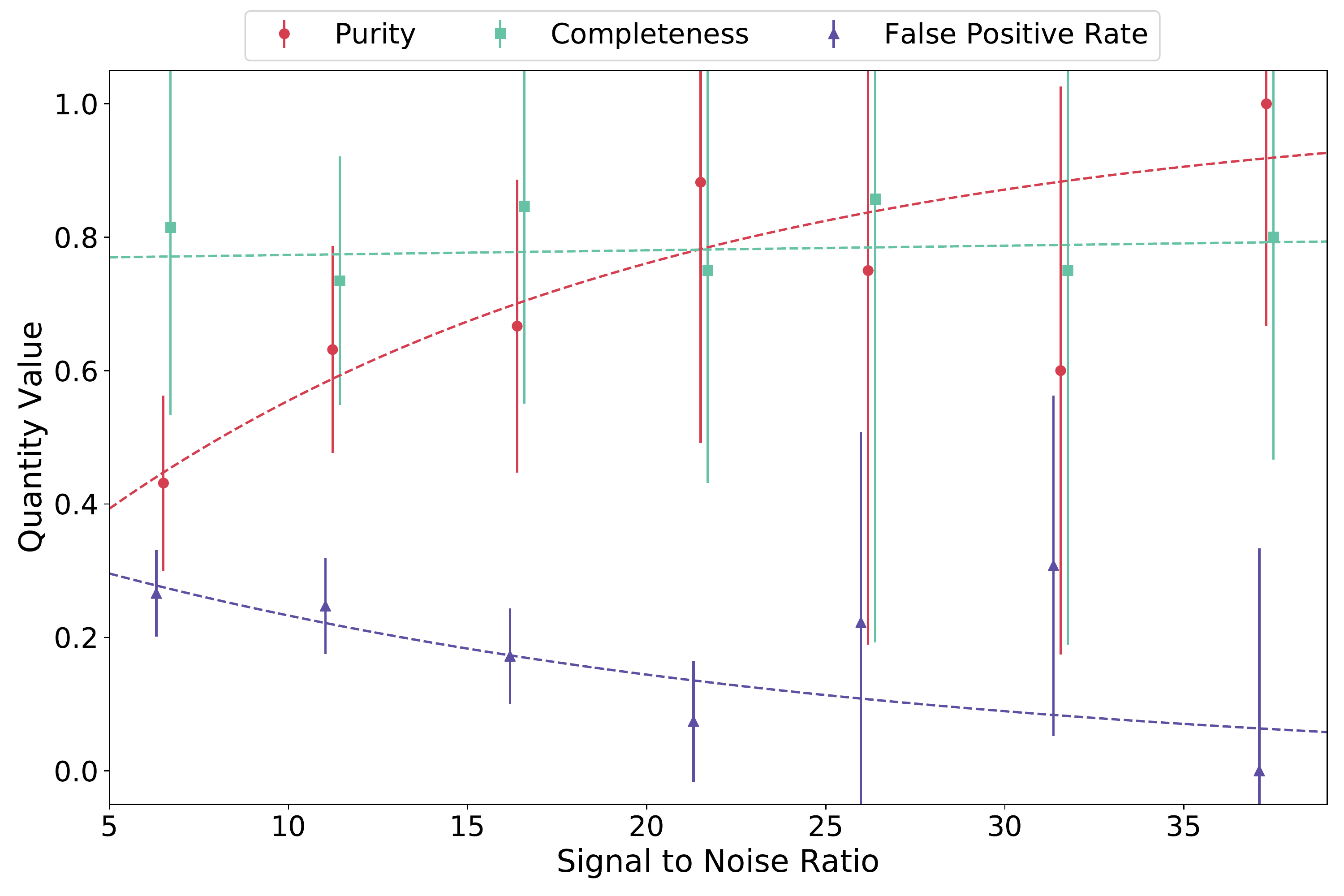}
    \caption{The purity, completeness, and false positive rate of the final output with probabilities greater than 0.5 considered a positive result. The points are determined using the mean value in each bin. The error bars become large on the edges of the apparent magnitude and signal to noise range due to the decreasing number of counts. Since images with SNR below 3.76 were removed during preprocessing, the SNR plot begins around 5.}
    \label{fig:mag_and_snr}
\end{figure*}

\subsection{Candidate Selection Function}
\label{sec:selection}

The selection function of our approach can also be determined using the purity, completeness, and false positive rate (FPR) of the output probabilities of the algorithm.
The probabilities can vary depending on the properties of the images such as the transient apparent magnitude and signal to noise ratio (SNR).
As the brightness of transients in images increases, the purity steadily increases to $93\%$ and the FPR decreases to less than $5\%$.
Similarly, for increasing SNR, the purity reaches values of $93\%$ and a FPR of $10\%$ as shown in Figure \ref{fig:mag_and_snr}.
These results demonstrate that our algorithm can effectively identify images with distinguishable features.
The completeness does not change significantly because the transient + host class is consistently given probabilities above 0.5 at a rate of about $80\%$ distributed over a large range of apparent magnitudes and SNR. 
For typical apparent magnitude and SNR ranges near the center of both plots from Figure \ref{fig:mag_and_snr}, we expect purity above 80\% for detections of real transients.

\subsection{Testing on Real Observations}
\label{sec:real_test}

Ultimately, the goal of this processing algorithm is to decrease the number of images requiring visual inspection. 
We can measure an increase in efficiency by calculating how many images fall above a 0.5 threshold from the output of the 2nd CNN and comparing to the number of images with \texttt{autoscan} probability above 0.7.
In reality, our approach is only meant to score the images based on their probability of being a host + transient, but we adopt a threshold to make comparisons to \texttt{autoscan}.
Values above these thresholds are used as standard indicators of a potential transient + host galaxy. 
After seeing improvements using only our method to filter potential candidates, we also applied a weighted combination of the \texttt{autoscan} score and our method to produce even better results in terms of the detections above a chosen threshold of 0.8.

The real datasets tested consist of samples of approximately 1,000 stamps from the total population of difference imaging detections.
To boost the representation of the host + transient class, part of each sample was collected by sampling objects with an \texttt{autoscan} score above 0.7 and part of each sample was collected by sampling objects randomly.
This sampling procedure produced a different distribution of 
\texttt{autoscan} scores in the samples than the full population, so we weight the reported detections to correct for that difference.
Essentially, we place the population and sample \texttt{autoscan} score probability distribution functions (PDFs) in bins of 0.05 and determine the factor required to scale the value of the sample PDF to the population PDF in each bin.
We also present the number of images in units of detections per square degree per night.
Thus, all test datasets are approximately on the same footing and can be directly compared.

The results from these calculations are shown in Table \ref{table:test_results}. 
The number of detections per square degree per night above the thresholds for our method are generally lower in most cases compared to \texttt{autoscan}. 
The combined method of calculating a probability seems to improve on our method across all tested datasets. 
Focusing specifically on the transient + host class, the number of images below our algorithm's threshold (column 6) and \texttt{autoscan} (column 5) demonstrates that both methods do not completely capture the desired images.
The significant decrease in the total number of detections using the combined method comes at the cost of fewer transient + host detections above the threshold.
The fraction of images passing the thresholds that are transient + host for our algorithm (column 9) and \texttt{autoscan} shows the purity of the detections that require inspection.
The combined method improves on the purity of detections for all of our tested datasets.
Many of the incorrectly passed images are not obvious transient due to the marginal quality of the difference image, so a fairly low purity is expected.

\begin{table*}
 \centering
 \begin{tabular}{|p{1.5cm}||p{1.4cm}|p{1.4cm}|p{1.4cm}||p{1.4cm}|p{1.4cm}|p{1.4cm}||p{1.4cm}|p{1.4cm}|p{1.4cm}|}
 \hline
 \multicolumn{10}{|c|}{Testing Results} \\
 \hline
 \multicolumn{1}{|c||}{} & \multicolumn{3}{>{\centering\arraybackslash}m{5cm}||}{Detections Above Threshold} & \multicolumn{3}{>{\centering\arraybackslash}m{5cm}||}{Transient + Host Detections Below Threshold} & \multicolumn{3}{>{\centering\arraybackslash}m{5cm}|}{Fraction Above Threshold that are Transient + Host Detections} \\
 \hline
 Dataset & \texttt{autoscan} & Our method & Combined & \texttt{autoscan} & Our method & Combined & \texttt{autoscan} & Our method & Combined \\
 \hline
 IC201114A & 398.712 & 77.648 & 51.304 & 3.732 & 98.616 & 91.793 & 0.437 & 0.536 & 0.952 \\
 IC171106A & 99.508 & 73.178 & 36.946 & 9.794 & 10.832 & 17.655 & 0.351 & 0.231 & 0.591 \\
 GW190814 & 152.368 & 86.072 & 43.047 & 7.664 & 6.335 & 12.460 & 0.080 & 0.150 & 0.170 \\
 IC190331A & 25.056 & 682.883 & 6.725 & 2.536 & 3.454 & 11.754 & 0.012 & 0.014 & 0.013 \\
 GW190728 & 84.029 & 358.448 & 29.255 & 0.185 & 23.494 & 25.816 & 0.010 & 0.044 & 0.011 \\
 \hline
 \end{tabular}
 \caption{The results from testing our method on real follow-up datasets compared to \texttt{autoscan} and the combined method. The values are given in units of detections per square degree per night. A detection was passed by each method if it was greater than an output probability threshold of 0.5 (our method), 0.7 (\texttt{autoscan}), and 0.8 (combined).}
 \label{table:test_results}
\end{table*}

%% file: discussion.tex
The development of our approach to do candidate counterpart identification using a convolutional neural network was motivated by the desire to improve the efficiency of multimessenger follow-ups. 
The efficacy of our algorithm and testing on real observations show improvements to the present method of image processing in the DESGW pipeline using \texttt{autoscan}.
The improvements decrease the number of images requiring visual inspection by researchers which means the identification of real transient + host objects can be faster.
Increased efficiency will be beneficial in the current era of multimessenger astronomy and even more so for future analyses using more advanced detectors with higher rates of events.
The implementation of our algorithm will improve the processing technique by making it more robust for future data, in turn, it will be easier to find useful and interesting objects.

The improved efficiency is reflected in the purity of the passed images between our method and \texttt{autoscan} as shown in columns 8 and 9 of Table \ref{table:test_results}. 
With generally larger fractions of passed transient + host images, fewer images need to be disregarded in the search for real objects.
\texttt{autoscan}'s lower purity is caused by overestimating the probability of being real with high values for other classes. 
The distribution of these high scores are shown in Figure \ref{fig:autoscan-compare} with similar median values for the transient + host, not obvious transient, and pre-existing point source classes.
Our method has lower certainty for what constitutes a real transient + host with smaller probabilities, but notably higher values than the other two classes meaning the probabilities can be used as improved indicators of real transients.
A threshold probability of 0.5 effectively achieves higher completeness and purity than an \texttt{autoscan} probability of 0.7.

Realtime follow-ups are dominated by false positives such as point sources and not obvious transients which slow down the identification of real optical counterparts in multimessenger searches.
The purity of the transient + host class above the thresholds is shown in columns 8-10 of Table \ref{table:test_results} with our method and the combined method outperforming \texttt{autoscan}.
Candidate identification can happen approximately 1.5 times faster using our method alone and 3.6 times faster in combinations with \texttt{autoscan} because there is a decreased number of images requiring inspection with improved purity and completeness of the transient + host class.
The benefits of higher efficiency will become even greater in the next era of gravitational wave detectors.
We expect a huge increase in the rate of events requiring triggered follow-ups which means faster responses will make real detections easier to find.
By integrating our tool into the DESGW pipeline, we have prepared DECam for the increased event rate which will lead to a higher probability of detecting the next multimessenger counterpart.

The deployment of our image processing algorithm will improve the efficiency and scalability of multimessenger counterpart searches with DECam.
The output probabilities from these detection methods can be used to set thresholds that filter the vast majority of false detections while maintaining the transient + host detections.
As a result, there will be a decreased number of images requiring visual inspection compared to \texttt{autoscan} and higher purity and completeness of datasets which demonstrates the efficacy of our method and the combined method.
These improvements will become more important during the next generation of multimessenger astronomy when more advanced detectors have higher event rates.

%% file: conclusion.tex
We developed an algorithm with the goal of improving multimessenger counterpart searches. 
Presently, these follow-up studies are plagued by high rates of false positive detections, primarily in the form of pre-existing point source objects already visible in the template image and not obviously real transient + host galaxy images.
The various classes of images have varying recognizable features that are visible in the pixel values of the images.
This visual aspect of classification motivated the use of convolutional neural networks in our algorithm to identify the real transients + host images from false cases.
Prior to reaching the CNNs, the images go through a series of preprocessing routines to simplify the classification and eliminate images that cause difficulties.
Each CNN has the same structure of 3 convolutional layers followed by a series of activation functions that decrease the output to an array of two probabilities corresponding to the positive and negative cases.
The CNNs differ by their training: the first network was trained to identify bad subtractions to be removed from the process, the second network was trained to identify transient + host images.
The images classified as not bad subtractions were saved and passed along to the second network where the images were given a probability of being a real transient + host.

The preprocessing removed other artifacts with a completeness and purity of 0.99. 
The first and second network were trained to accuracies of 92\% and 72\% respectively. 
The first network's ROC curve shown in the left panel of Figure \ref{fig:network1_performance} was calculated with an AUC of 0.982 and the operating threshold was used to find a 92\% true positive rate.
The second network's success is shown in Figure \ref{fig:autoscan-compare}. The majority of the transient + host class received probabilities above a 0.5 threshold whereas the most common types of false positives (no obvious transient and pre-existing point source) fall below this value.
The output probabilities of the second network compared to existing artifact detection software demonstrate our algorithm's ability to distinguish true positives from false positives.
We tested the final end-to-end algorithm on five unseen real follow-up datasets and demonstrated improvements with a decreased number of images requiring visual inspection using our method.
We also created a weighted combination of our method and \texttt{autoscan} that further improved our results and decreased the number of images requiring inspection by 3.6.

The results of the various tests and comparisons demonstrate the success of our algorithm at processing sets of search, template, and difference images such that real transient + host galaxies can be identified more efficiently. 
Such improvements over the current method built into the DESGW pipeline using \texttt{autoscan} will become especially beneficial during the next era of multimessenger astronomy when more advanced gravitational wave detectors go online.
Decreasing the amount of required human inspection will expedite the search and the integration of our method into the DESGW pipeline will increase the probability of detecting a multimessenger counterpart.

%% file: acknowledgements.tex
R. Morgan thanks the LSSTC Data Science Fellowship Program, which is funded by LSSTC, NSF Cybertraining Grant \#1829740, the Brinson Foundation, and the Moore Foundation; his participation in the program has benefited this work. 

This material is based upon work supported by the National Science Foundation Graduate Research Fellowship Program under Grant No. 1744555. Any opinions, findings, and conclusions or recommendations expressed in this material are those of the author(s) and do not necessarily reflect the views of the National Science Foundation.

Funding for the DES Projects has been provided by the U.S. Department of Energy, the U.S. National Science Foundation, the Ministry of Science and Education of Spain, 
the Science and Technology Facilities Council of the United Kingdom, the Higher Education Funding Council for England, the National Center for Supercomputing 
Applications at the University of Illinois at Urbana-Champaign, the Kavli Institute of Cosmological Physics at the University of Chicago, 
the Center for Cosmology and Astro-Particle Physics at the Ohio State University,
the Mitchell Institute for Fundamental Physics and Astronomy at Texas A\&M University, Financiadora de Estudos e Projetos, 
Funda{\c c}{\~a}o Carlos Chagas Filho de Amparo {\`a} Pesquisa do Estado do Rio de Janeiro, Conselho Nacional de Desenvolvimento Cient{\'i}fico e Tecnol{\'o}gico and 
the Minist{\'e}rio da Ci{\^e}ncia, Tecnologia e Inova{\c c}{\~a}o, the Deutsche Forschungsgemeinschaft and the Collaborating Institutions in the Dark Energy Survey. 

The Collaborating Institutions are Argonne National Laboratory, the University of California at Santa Cruz, the University of Cambridge, Centro de Investigaciones Energ{\'e}ticas, 
Medioambientales y Tecnol{\'o}gicas-Madrid, the University of Chicago, University College London, the DES-Brazil Consortium, the University of Edinburgh, 
the Eidgen{\"o}ssische Technische Hochschule (ETH) Z{\"u}rich, 
Fermi National Accelerator Laboratory, the University of Illinois at Urbana-Champaign, the Institut de Ci{\`e}ncies de l'Espai (IEEC/CSIC), 
the Institut de F{\'i}sica d'Altes Energies, Lawrence Berkeley National Laboratory, the Ludwig-Maximilians Universit{\"a}t M{\"u}nchen and the associated Excellence Cluster Universe, 
the University of Michigan, NFS's NOIRLab, the University of Nottingham, The Ohio State University, the University of Pennsylvania, the University of Portsmouth, 
SLAC National Accelerator Laboratory, Stanford University, the University of Sussex, Texas A\&M University, and the OzDES Membership Consortium.

Based in part on observations at Cerro Tololo Inter-American Observatory at NSF’s NOIRLab (NOIRLab Prop. ID 2012B-0001; PI: J. Frieman), which is managed by the Association of Universities for Research in Astronomy (AURA) under a cooperative agreement with the National Science Foundation.

The DES data management system is supported by the National Science Foundation under Grant Numbers AST-1138766 and AST-1536171.
The DES participants from Spanish institutions are partially supported by MICINN under grants ESP2017-89838, PGC2018-094773, PGC2018-102021, SEV-2016-0588, SEV-2016-0597, and MDM-2015-0509, some of which include ERDF funds from the European Union. IFAE is partially funded by the CERCA program of the Generalitat de Catalunya.
Research leading to these results has received funding from the European Research
Council under the European Union's Seventh Framework Program (FP7/2007-2013) including ERC grant agreements 240672, 291329, and 306478.
We  acknowledge support from the Brazilian Instituto Nacional de Ci\^encia
e Tecnologia (INCT) e-Universe (CNPq grant 465376/2014-2).

This paper has gone through internal review by the DES collaboration.
This manuscript has been authored by Fermi Research Alliance, LLC under Contract No. DE-AC02-07CH11359 with the U.S. Department of Energy, Office of Science, Office of High Energy Physics.

%% file: appendix.tex
All of the labels from \texttt{ArtifactSpy} were completely determined by human inspection.
While this ensured that the CNNs would be tested against verifiable images, the subjective labels could not be expected to be 100\% consistent. 
Even with a standard set of defining characteristics of each class, many types of images were difficult to always identify as a certain class, especially with a team of people applying labels. 
Preliminary tests of the CNN performance indicated a significant amount of misclassified bad subtractions. 
Closer inspection of the incorrect predictions showed that human error may have led to incorrect labels.

\begin{figure*}
    \centering
    \includegraphics[trim={1.5cm 3cm 2cm 1cm},clip,width=\linewidth]{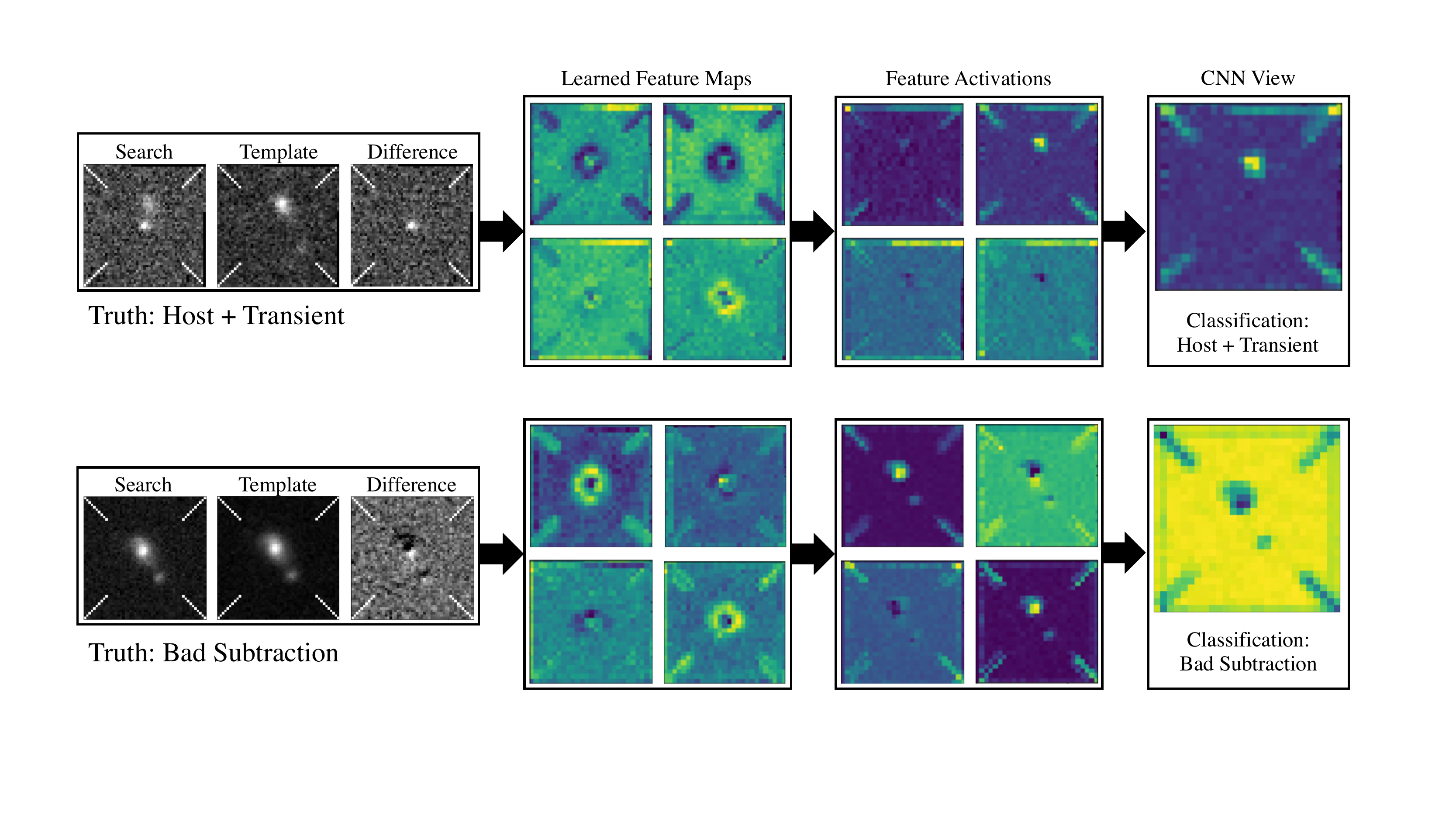}
    \caption{A visualization of how sets of difference images are interpreted by our networks. After training, the CNNs have learned what to look for in the images to make a classification. This information is stored in the gradients of the final convolutional layer of the network and is shown by the ``Learned Feature Maps'' column. After convolving the learned feature map with the image, the output is passed through a ReLU activation function to produce the class activation map shown as the ``CNN View'' column. The learned feature maps demonstrate the network is looking for point-like objects in the center of the images, elongated and off-center objects, and dipole-like regions. The bright spots in the ``CNN View'' of the host + transient image and inversely the dark spots for the bad subtraction indicate where those features are located.}
    \label{fig:gradcam}
\end{figure*}

To inspect misclassified images, we employed a technique called Gradient-weighted Class Activation Mapping \citep[Grad-CAM;][]{gradcam} to highlight regions on the images with the highest contributions towards the CNN's classification decision. 
Convolutional layers retain spatial information that informs classifications in the fully-connected layers.
Grad-CAM saves the gradients of each class with respect to the activation maps of the last convolutional layer to produce a localization map, essentially enabling us to inspect what the network sees right before the images are classified. 
The resulting heatmap of the difference images showed that the network was identifying the same features we had been using in our by-eye classifications; the final classification was only incorrect because the initial label was incorrect.
The images determined to have incorrect initial labels were mostly subtle bad subtractions or transients with small amounts of host separation, making them hard for the team of labelers to agree on.
With the Grad-CAM technique serving as a lens, we were able to identify when the neural network was making a correct, but wrong due to the initial label of an image, classification and corrected the initial labels in the difference imaging data.
An example of this process is shown in Figure \ref{fig:gradcam}.

